\documentclass[runningheads]{llncs}
\usepackage[T1]{fontenc}

\usepackage{graphicx}

\usepackage{hyperref}
\usepackage{color}

\usepackage{amsmath,amssymb,amsfonts}
\usepackage{xcolor} 
\usepackage{subcaption}
\usepackage{orcidlink}

\begin{document}
\title{Towards Track Height Control in Additive Manufacturing on Porous Substrates: An Approach for System Identification and Controller Selection}
\titlerunning{Towards Track Height Control in AM on Porous Substrates}

\author{Amelie Schatull\thanks{Corresponding author.}\inst{1}\orcidlink{0009-0001-5326-9076} \and
Torge Kolditz\inst{2}\orcidlink{0000-0003-0646-9363} \and
Simon Kleinschmidt\inst{2}\orcidlink{0009-0009-9141-5132} \and
Annika Raatz\inst{2}\orcidlink{0000-0002-1697-1907} \and
J\"org Seewig\inst{1}\orcidlink{0000-0002-1420-1597}}
\authorrunning{A. Schatull et al.}

\institute{Institute for Measurement and Sensor Technology, RPTU University Kaiserslautern-Landau, 67663, Kaiserslautern, Germany \\
\email{a.schatull@mv.rptu.de} \and
Institute of Assembly Technology and Robotics, Leibniz University Hannover, 30823, Garbsen, Germany}

\maketitle              
\begin{abstract}
Additive manufacturing is transforming the creation of materials and parts, enabling the design and manufacturing of multifunctional parts with tailored properties. One compelling example is the combination of a porous core and a non-porous surface, which offers significant advantages in terms of weight reduction, material efficiency, strength and wear resistance. Directed Energy Deposition (DED) processes enable the transition from porous core to non-porous part surface by closing the pores.
However, the surface-layer of porous substrates presents a substantial challenge to achieving consistent track heights during DED, directly impacting the final part geometry and performance. Therefore, precise control of the deposition process is crucial for realising the full potential of the multifunctional parts. 
In order to address this challenge, this work presents a conceptual approach for track height control during Laser-based Directed Energy Deposition (DED-LB) on porous substrates.
First, the DED-LB process and the associated measurement setup are explained, along with a discussion of their limitations. The focus then shifts to the control problem, beginning with system identification to characterise the process dynamics. Subsequently, a selection of control strategies is investigated, as well as their compatibility with the process. This research provides a foundational step toward reliable additive manufacturing on porous substrates, and thus toward producing multifunctional parts.

\end{abstract}

\section{Introduction}\label{intr}
Within the last years, the interest in additive manufacturing (AM) has significantly increased \cite{Alfaify2020} as it provides several advantages, such as near-net shape, reduced waste and manufacturing of freeform parts \cite{Tapia2014}.
Moreover, especially DED processes, such as DED-LB or wire arc additive manufacturing (WAAM), enable the manufacturing of multi-functional parts having a porous core and a non-porous durable surface. 
The porous core provides not only weight reduction within the part, but can also be used to tailor thermal properties and energy absorption locally \cite{Rabiei2024}. 

\begin{figure}[]
    \centering
    \includegraphics[width=0.8\linewidth]{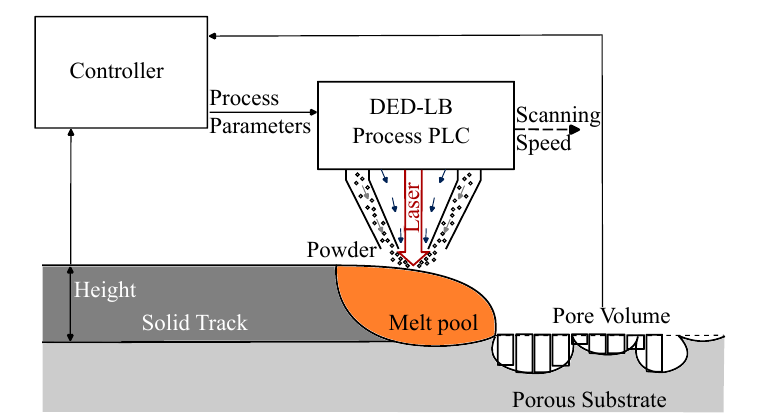}
    \caption{A schematic representation of the DED-LB process on a porous substrate with the control structure: The pore volume and the track height are the control inputs, while the new process parameters are fed into the PLC to set the laser power and scanning speed.}
    \label{fig:ControlSetup}
\end{figure}

However, manufacturing parts that meet quality standards remains challenging for DED, particularly when building on porous surfaces. Consequently, a key challenge when working with porous substrates are geometrical distortions along the deposited tracks as they traverse multiple layers.
In this work, we address the challenge employing a method for different control approaches, with DED-LB as an example process.
In figure \ref{fig:ControlSetup} the control problem is visualised: when the melt pool and track manufactured by the DED-LB process meet surface pores, a controller must compensate for the disturbances in height. Therefore, the controller adjusts process parameters such as scanning speed based on the measured pore volume or track height. Here, we investigate the concepts of feedforward control,  feedback control and lookup table control, with regard to mitigating geometrical distortions in the context of DED-LB on porous substrates. 

The presented paper is structured as follows: Section \ref{sec:DED} describes the DED-LB process, related work, the measurement setup and the overall concept. Section \ref{sec:Identification} identifies different approaches to generate identification data, while section \ref{sec:Control} discusses and compares different control methods. Section \ref{sec:conclusion} summarises the investigation.

\section{Related Work}\label{sec:DED}
DED is an AM process which uses metal powder or wire as feedstock and a heat source, such as a laser, electron beam or electric arc, to manufacture parts \cite{Gibson2020}. In this study, we focus on the laser-based process using AISI 316L stainless steel powder as the feedstock. The functionality of the process is visualised in figure \ref{fig:ControlSetup}: A laser heat source melts the substrate, forming a so called melt pool on the surface \cite{Borovkov2021}. Meanwhile a carrier gas flow transports the powder particles, which are protected from oxidation by a shielding gas, to the melt pool, where they fuse with the molten substrate material \cite{Borovkov2021}. Moving the laser beam and powder stream along the substrate surface at an appropriate scanning speed then forms a melt track that solidifies as it cools \cite{Bayat2021}. 
In this way parts can be manufactured in a layer by layer manner, with each layer adding to the overall structure. This track- and layer-wise construction method provides the opportunity to create freeform structures \cite{Gibson2020} and manufacture on porous substrates.

In order to influence the track geometry on a porous substrate, different machine parameters can be varied.
Thus, controllers in DED-LB often adapt parameters such as scanning speed $v$, laser power $P$ and powder mass rate $m_\text{p}$ \cite{Cao2015}.
Within literature, different controllers have been investigated in the context of non-porous substrates.
Cao and Ayalew \cite{Cao2019} developed a multi-input-multi-output (MIMO) model to address geometric control in laser-aided powder deposition. The parameter identification was conducted via finite element method (FEM) simulations, and the resulting nonlinear model predictive controller (MPC) combined feedforward and feedback control based on local linearisation. Consequently, the MIMO model improved surface finish and accurately represented temperature behaviour.
Garmendia et al. \cite{Garmendia2018} proposed an offline, layer-wise control strategy combined with online measurements using a structured light-based 3D scanner. Thus, height deviations in laser metal deposition were reduced by adding or deleting layers.
To control the clad height, Fathi et al. \cite{Fathi2007} adapted the scanning speed through a PID controller based on a Hammerstein model. Combined with a feedforward module, the experimental results yielded faster response times and reduced overshoot.
In 2019, Dillk\"otter and M\"onnigmann \cite{Dillkotter2019} investigated a model-based discrete-time feedforward controller in direct metal deposition. By adjusting the laser power, the melt pool width was controlled, resulting in good performance compared to a benchmark.

In this study, we focus on three different controller types: a feedforward controller, a feedback controller and a lookup table.
As all those controllers require geometry data of the manufactured tracks, those have to be measured. 
In this instance, the track dimensions are determined by means of a laserline triangulation sensor, which measures both the track height and width.
The sensor used is the ScanControl3000-25/BL by Micro-Epsilon \cite{Micro-Epsilon2023}, which has a wavelength of $405$\,nm and a 2048$\times$1088 sensor matrix covering a 25\,mm$\times$15\,mm field of view.
The same sensor is also used to enhance the pore geometries of the porous substrate.
Based on this information, the feedforward controller, for example, calculates the required process parameters in advance.

\begin{figure}
    \centering
    \includegraphics[width=0.5\linewidth]{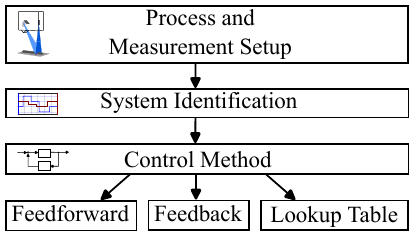}
    \caption{Steps of the investigated conceptual approach that aim to identify a control method for porous substrates.}
    \label{fig:Concept}
\end{figure}

Moreover, the three investigated controllers all have in common, that first a system identification is needed in order to identify the control model parameters. 
Thus, this is the next step after definition of the measurement setup. The overall approach is depicted in figure \ref{fig:Concept}. It describes not only the steps to build a DED-LB control method, but also the steps of this paper.

\section{System Identification}\label{sec:Identification}
System identification enables the adaption of a process model towards the specific process being investigated. Thus, based on this data, the empirical parameters of model-based control approaches can be identified or the data base of tabular models \cite{Bohn2016} can be built.
Therefore, system identification involves varying the process parameters in a parameter study to enhance the system's dynamics and influence of parameters. 

Conventionally, parameter identification in AM involves manufacturing single tracks, each with a different parameter configuration.
For a classical lookup table this procedure is sufficient.
In contrast, model-based approaches require not only the stationary track values but also the dynamical behaviour of the system. 
In order to enhance this additional information without the necessity for additional tracks, it seems beneficial to vary process parameters along a single track.
This captures the process reaction to parameter changes, as well as the stationary track values resulting from those parameters.

\begin{figure}[ht]
\centering
\begin{subfigure}{0.49\textwidth}
    \includegraphics[width=\linewidth]{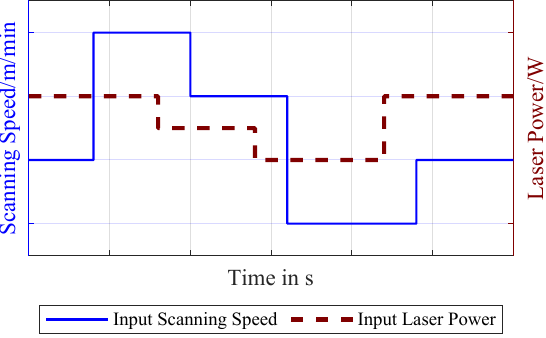}
    \caption{}
    \label{fig:Input}
\end{subfigure}
\hfill
\begin{subfigure}{0.49\textwidth}
    \includegraphics[width=\textwidth]{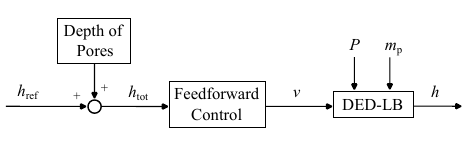}
    \caption{}
    \label{fig:feedforward}
\end{subfigure}
\hfill
\begin{subfigure}{0.49\textwidth}
    \includegraphics[width=\textwidth]{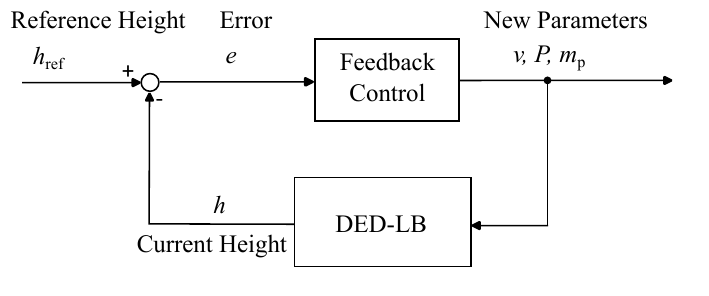}
    \caption{}
    \label{fig:feedback}
\end{subfigure}
\hfill
\begin{subfigure}{0.49\textwidth}
    \includegraphics[width=\textwidth]{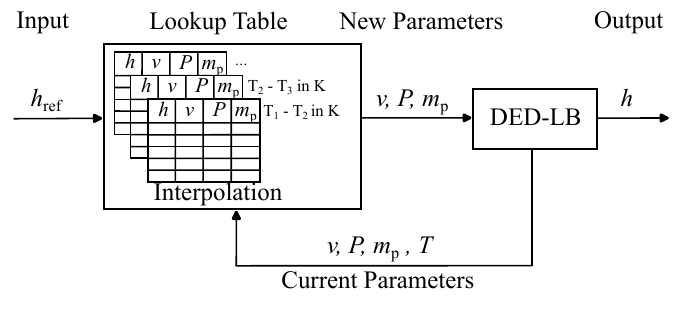}
    \caption{}
    \label{fig:lookup}
\end{subfigure}
\caption{(a) step-wise input variation of laser power and scanning speed over time to enhance the system's dynamical and steady state behaviour; (b) feedforward height control in DED-LB, reproduced from \cite{Schatull2026} (Fig.2), licensed under CC BY 4.0 (\url{https://creativecommons.org/licenses/by/4.0/}), varying scanning speed based on measured pore depth; (c) feedback height control in DED-LB minimising error between current height and desired reference height; (d) lookup table height control in DED-LB estimating process parameters through interpolation}
\end{figure}

Thus, by varying the system's input in a step-wise manner, as illustrated in figure \ref{fig:Input}, the dynamics of the process can be enhanced and used in the process model. 
This approach seems promising as it has already been used within literature.
Liu et al. \cite{Liu2019} developed a transfer function model and identified the empirical parameters by varying the laser power input while keeping the other process parameters constant. The step-wise change of laser power input provokes a reaction in the track behaviour such as higher melt pool width at higher laser power \cite{Liu2019}. Furthermore, by altering the laser power along a single track, they were able to capture the system's dynamic response information in addition to the steady state track width \cite{Liu2019}.
Cao and Ayalew \cite{Cao2019} used a similar input method, but in addition to altering the laser power, they also varied the scanning speed  in a step-wise manner to influence the track height. 
The resulting track output was then used to estimate the empirical parameters of their multi-input-multi-output (MIMO) model via an objective function \cite{Cao2019}.
The first system identification with the process and sensor setup under investigation has already been accomplished \cite{Schatull2025}.

Additionally investigating non-step inputs, such as sinusoidal signals, would reveal dynamics and potential non-linearities that step tests might miss. Moreover, linear ramps could smoothen the transition between input changes, thus providing a first basis on how to create the controller input for smooth tracks.
The so generated system identification data provides the foundation for estimating empirical parameters for the model approaches discussed in the following.

\section{Control Methods}\label{sec:Control}
In the following investigation, three control approaches are investigated and compared, namely feedforward control, feedback control and the lookup table method. The advantages and disadvantages of each are considered, especially in terms of production aspects.
In the field of control methods, our focus is mainly on model-based control, since the investigated process is physically complex. Those control models can be categorised into three groups: black-box, grey-box and white-box \cite{Bohn2016}. While white-box models are based solely on physical knowledge and equations, black-box models are based solely on data generated via experiments or simulations \cite{Bohn2016}. Grey-box models, however, combine the advantages of both: they describe the system using physical equations combined with empirical parameters that need to be identified \cite{Bohn2016}. Thus, the model can be adapted to the process via parameters while also taking into account the underlying physics.
Therefore, it is advisable to compare the different types of models, their respective advantages and disadvantages, in order to select an appropriate one for the investigated process. 
In the following, a comparison will be drawn between feedforward, feedback and lookup table control method.

\paragraph{Feedforward Control}
The use of the feedforward control principle within AM is already discussed within literature. Dillkötter and Mönnigmann \cite{Dillkotter2019}, for example, propose a feedforward controller based on the Rosenthal equation. This controller adjusts the laser power in order to enhance a constant melt pool width \cite{Dillkotter2019}.

In this work, we aim to overcome height-disturbances in DED-LB caused by a porous substrate, and one approach is a feedforward controller \cite{Garmendia2018}. 
As can be seen in figure \ref{fig:feedforward} the desired reference height $h_\text{ref}$ is combined with the offline pre-measured pore depth. The feedforward controller then adjusts the scanning speed $v$ accordingly to compensate for the pores. This adaption takes place offline and creates the machine G-Code in advance. Next, tracks are manufactured on porous substrates via the DED-LB process. Thus, the reaction time of the controller does not affect the process, as the G-Code is pre-computed. Moreover, there is no need to access the DED-LB process online or make the laser line triangulation sensor heat resistant.
However, the feedforward controller is limited to the pre-measured pore depths and is unable to react online to variations in track height during the process, as the information only flows in one direction \cite{Garmendia2018}.

\paragraph{Feedback Control}
Unlike a feedforward controller, a feedback controller can react to changes in track height in real time. The general method is depicted in figure \ref{fig:feedback}: A desired reference height $h_\text{ref}$ is compared to the current track height $h$, for example measured by the laser line triangulation sensor. The resulting difference, also known as the error $e$ \cite{Bohn2016}, is given as input to the controller which then calculates the new process parameters.
In this way, the difference between $h_\text{ref}$ and $h$ can be minimised online, enabling the feedback controller to react to changes in track height, width or temperature during the process. However, the feedback controller has to be implemented as real-time capable.
Therefore, the computing hardware must also meet those requirements, and in order to improve response times the code needs to be optimised.
Additionally, the DED-LB machine needs to provide open communication interfaces, thereby enabling the controller to transmit new process inputs to the PLC during the process.
This can be achieved in different ways, for example with the controller running on an external IPC, processing measurement data and transmitting the new PLC inputs.
For this transmission a communication protocol, such as PROFINET for a Siemens PLC \cite{Kleines2008}, is needed. 
Moreover, the laser line triangulation sensor requires a trigger signal from the PLC to initiate the measurement while logging the time signal. 
This allows the movement and position of the machine axis to be associated with the measurement values acquired by the sensor. 
This trigger signal must be configured not only in the G-Code and sensor software, but also in the PLC interface hardware. 
Nevertheless, we can use the knowledge of the pre-measured pores to simplify the control problem for the feedback controller by additionally using a feedforward controller. This method is already used by Fathi et al. \cite{Fathi2007} and according to their results, combining a feedforward and a feedback controller results in lower response times and smaller overshoot as the feedforward controller reduces disturbances in advance.

\paragraph{Lookup Table}
Another method of achieving real-time control is the use of lookup tables. Within this tabular control method \cite{Bohn2016}, different parameter configurations and their outputs are pre-calculated or pre-measured and then interpolated during the control process \cite{Coffey2011}.
In the context of the DED-LB process, the resulting track heights, widths and temperatures for different combinations of input parameters $v$, $P$ and $m_\text{p}$ have to be determined experimentally, for example through a design of experiment process (DoE).
Based on these results, the optimal parameter configuration can be rapidly determined by linear interpolation during the process. Figure \ref{fig:lookup} illustrates this control setup, which adapts process parameters in order to obtain the desired track height $h_\text{ref}$.
Considering the current process temperature $T$ and set-point ($v$, $P$, $m_\text{p}$), the lookup table data is searched and linear interpolated for a fitting parameter constellation to reach the desired track height $h_\text{ref}$.
The so determined new process parameters are entered as input into the process and the track height responds, which can be seen in figure \ref{fig:lookup}.
Even if the lookup table is quick and simple to incorporate into an existing setup, this method relies primarily on the inherent data.
The more data points and parameter combinations are measured in advance, the closer the linear interpolation approximates the desired parameter combination.
Thus, the DoE is essential for the lookup table controller results and restricts the controller range, as extrapolation is not advisable as the laser could destroy the machine.
Moreover, a change in setup, for example by switching to a different nozzle, requires many new experiments to acquire the necessary data.
Applying a process model similar to that employed for the feedback controller could enable lookup table inputs to be estimated, thereby reducing the number of experiments required.
Furthermore, using Gaussian processes instead of linear interpolation could improve the results, as they consider uncertainty.\\

With regard to the investigated DED-LB process and our individual process setup, the feedforward controller seems like a suitable first approach, as no adaptation of communication interfaces is required.
This way, a first control on porous substrates is possible, and the resulting tracks offer insight into process-specific challenges, such as material fusion or reflectivity.
Based on these initial controller approaches, it will be easier to develop a feedback controller while the communication interfaces are established and the sensor is made heat-resistant. 
As explained previously, combining a feedback controller with a feedforward controller may improve response times.

\section{Conclusion}\label{sec:conclusion}
The presented study analyses control approaches of track height in order to compensate for defects caused by porous substrates, with DED-LB as an example process.
The conceptual approach presented can be summarised as follows:
First, the system identification data needs to be enhanced, either through simulation or experimentally, for example using a laser line triangulation sensor.
Therefore, a step-wise changing in input parameters along a single track is recommended to enhance stationary and dynamic system information.
Next, a control model has to be chosen accordingly to the individual system requirements. In this study, we compared three approaches:
\begin{itemize}
    \item Feedforward Control: Following system identification, the machine G-Code is pre-calculated and run without the need for open communication interfaces, that would enable reading and writing during the process. However, an online reaction is not possible.
    \item Feedback Control: This method enables online control and thus also reaction to temperature, while requiring open communication interfaces by the DED-LB machine.
    \item Lookup Table: The model depends on the quality and range of measured data points. Thus, various experiments or simulations are required, while the linear interpolation is fast and real-time capable.
\end{itemize}
For the DED-LB process example under investigation, feedforward is the initial approach, as no adaptation of the machine setup is required.
The so enhanced knowledge of the process facilitates the development of a feedback controller in the next step.
Thus, the investigated control methods offer promising approaches to building tracks on porous substrates. Providing different advantages for different DED process setups, this approach is a first step toward reliable additive manufacturing on porous substrates, which is crucial for multifunctional parts having a porous core to reduce weight.

\begin{credits}
\subsubsection{\ackname} Funded by the Deutsche Forschungsgemeinschaft (DFG, German Research Foundation) – Project-ID 511263698 – TRR 375.

\subsubsection{\discintname}
The authors have no competing interests to declare that are relevant to the content of this article.
\end{credits}

\bibliographystyle{splncs04}
\bibliography{sources}

\begin{thebibliography}{10}
\providecommand{\url}[1]{\texttt{#1}}
\providecommand{\urlprefix}{URL }
\providecommand{\doi}[1]{https://doi.org/#1}

\bibitem{Alfaify2020}
Alfaify, A., Saleh, M., Abdullah, F.M., Al-Ahmari, A.M.: Design for {Additive}
  {Manufacturing}: {A} {Systematic} {Review}. Sustainability  \textbf{12}(19),
  ~7936 (Sep 2020). \doi{10.3390/su12197936},
  \url{https://www.mdpi.com/2071-1050/12/19/7936}

\bibitem{Bayat2021}
Bayat, M., Dong, W., Thorborg, J., To, A.C., Hattel, J.H.: A review of
  multi-scale and multi-physics simulations of metal additive manufacturing
  processes with focus on modeling strategies. Additive Manufacturing
  \textbf{47},  102278 (Nov 2021). \doi{10.1016/j.addma.2021.102278},
  \url{https://linkinghub.elsevier.com/retrieve/pii/S2214860421004383}

\bibitem{Bohn2016}
Bohn, C., Unbehauen, H.: Identifikation dynamischer {Systeme}. Springer
  Fachmedien Wiesbaden, Wiesbaden (2016). \doi{10.1007/978-3-8348-2197-3},
  \url{http://link.springer.com/10.1007/978-3-8348-2197-3}

\bibitem{Borovkov2021}
Borovkov, H., De~La~Yedra, A.G., Zurutuza, X., Angulo, X., Alvarez, P.,
  Pereira, J.C., Cortes, F.: In-{Line} {Height} {Measurement} {Technique} for
  {Directed} {Energy} {Deposition} {Processes}. Journal of Manufacturing and
  Materials Processing  \textbf{5}(3), ~85 (Aug 2021).
  \doi{10.3390/jmmp5030085}, \url{https://www.mdpi.com/2504-4494/5/3/85}

\bibitem{Cao2015}
Cao, X., Ayalew, B.: Control-oriented {MIMO} modeling of laser-aided powder
  deposition processes. In: 2015 {American} {Control} {Conference} ({ACC}). pp.
  3637--3642. IEEE, Chicago, IL, USA (Jul 2015).
  \doi{10.1109/ACC.2015.7171895},
  \url{http://ieeexplore.ieee.org/document/7171895/}

\bibitem{Cao2019}
Cao, X., Ayalew, B.: Robust multivariable predictive control for laser-aided
  powder deposition processes. Journal of the Franklin Institute
  \textbf{356}(5),  2505--2529 (Mar 2019).
  \doi{10.1016/j.jfranklin.2018.12.015},
  \url{https://linkinghub.elsevier.com/retrieve/pii/S0016003219300298}

\bibitem{Coffey2011}
Coffey, B.: Using {Building} {Simulation} and {Optimization} to {Calculate}
  {Lookup} {Tables} for {Control}. Ph.D. thesis (2011),
  \url{https://api.semanticscholar.org/CorpusID:60625939}

\bibitem{Dillkotter2019}
Dillkötter, D., Mönnigmann, M.: Design of a model based feedforward
  controller for additive manufacturing by laser metal deposition. In: 2019
  18th {European} {Control} {Conference} ({ECC}). pp. 3842--3847. Napoli (2019)

\bibitem{Fathi2007}
Fathi, A., Khajepour, A., Toyserkani, E., Durali, M.: Clad height control in
  laser solid freeform fabrication using a feedforward {PID} controller. The
  International Journal of Advanced Manufacturing Technology  \textbf{35}(3-4),
   280--292 (Dec 2007). \doi{10.1007/s00170-006-0721-1},
  \url{http://link.springer.com/10.1007/s00170-006-0721-1}

\bibitem{Garmendia2018}
Garmendia, I., Leunda, J., Pujana, J., Lamikiz, A.: In-process height control
  during laser metal deposition based on structured light {3D} scanning.
  Procedia CIRP  \textbf{68},  375--380 (2018).
  \doi{10.1016/j.procir.2017.12.098},
  \url{https://linkinghub.elsevier.com/retrieve/pii/S2212827117310399}

\bibitem{Gibson2020}
Gibson, B., Bandari, Y., Richardson, B., Henry, W., Vetland, E., Sundermann,
  T., Love, L.: Melt pool size control through multiple closed-loop modalities
  in laser-wire directed energy deposition of {Ti}-{6Al}-{4V}. Additive
  Manufacturing  \textbf{32},  100993 (Mar 2020).
  \doi{10.1016/j.addma.2019.100993},
  \url{https://linkinghub.elsevier.com/retrieve/pii/S2214860419317488}

\bibitem{Kleines2008}
Kleines, H., Detert, S., Drochner, M., Suxdorf, F.: Performance {Aspects} of
  {PROFINET} {IO}. IEEE Transactions on Nuclear Science  \textbf{55}(1),
  290--294 (2008). \doi{10.1109/TNS.2007.914032},
  \url{http://ieeexplore.ieee.org/document/4448497/}

\bibitem{Liu2019}
Liu, Y., Wang, L., Brandt, M.: Model predictive control of laser metal
  deposition. The International Journal of Advanced Manufacturing Technology
  \textbf{105}(1-4),  1055--1067 (Nov 2019). \doi{10.1007/s00170-019-04279-9},
  \url{http://link.springer.com/10.1007/s00170-019-04279-9}

\bibitem{Micro-Epsilon2023}
Micro-Epsilon: {scanCONTROL} {Configuration} {Tools} (2023),
  \url{https://www.micro-epsilon.de}

\bibitem{Rabiei2024}
Rabiei, A., Cance, J., Chacko, Z.: A {Study} on {Welding} of {Porous} {Metals}
  and {Metallic} {Foams}. Advanced Engineering Materials  \textbf{26}(4),
  2301430 (Feb 2024). \doi{10.1002/adem.202301430},
  \url{https://advanced.onlinelibrary.wiley.com/doi/10.1002/adem.202301430}

\bibitem{Schatull2025}
Schatull, A., Kleinschmidt, S., Platz, J., Aurich, J.C., Raatz, A., Seewig, J.:
  Control-oriented system identification methods in additive manufacturing: a
  comparison. Production Engineering  \textbf{20}(1), ~28 (Feb 2026).
  \doi{10.1007/s11740-025-01387-2},
  \url{https://link.springer.com/10.1007/s11740-025-01387-2}

\bibitem{Schatull2026}
Schatull, A., Platz, J., Bachert, L., Kleinschmidt, S., Raatz, A., Aurich,
  J.C., Seewig, J.: Feed rate modulation for filling macroscopic surface pores
  in directed energy deposition: An experimental evaluation using boreholes as
  representative geometries. Progress in Additive Manufacturing  (2026).
  \doi{10.1007/s40964-026-01902-6}

\bibitem{Tapia2014}
Tapia, G., Elwany, A.: A {Review} on {Process} {Monitoring} and {Control} in
  {Metal}-{Based} {Additive} {Manufacturing}. Journal of Manufacturing Science
  and Engineering  \textbf{136}(6),  060801 (Dec 2014).
  \doi{10.1115/1.4028540},
  \url{https://asmedigitalcollection.asme.org/manufacturingscience/article/doi/10.1115/1.4028540/377521/A-Review-on-Process-Monitoring-and-Control-in}

\end{thebibliography}

\end{document}